\documentstyle[aps,epsf,epsfig]{revtex}
\newcommand{\dem}{\mbox{$\frac{1}{2}$}}

\begin{document}
\title{{\bf Thermodynamics of a mixed quantum-classical Heisenberg model in two
dimensions}}
\author{J. Leandri, Y. Leroyer, S.V. Meshkov and Y. Meurdesoif}
\address{Centre de Physique Th\'eorique et de Mod\'elisation de Bordeaux\\
Universit\'e Bordeaux I, CNRS, Unit\'e Associ\'ee 1537\\
19 rue du Solarium, 33174 Gradignan Cedex, France
\\{\rm and}}
\author{O. Kahn, B. Mombelli and D. Price}
\address{Laboratoire des Sciences Mol'culaires\\
Institut de Chimie de la Mati\`ere Condens\'ee de Bordeaux\\
CNRS, UPR 9048\\
Avenue Albert Schweitzer, 33608 Pessac Cedex, France}
\maketitle
\thispagestyle{empty}
\begin{abstract}
We study the planar antiferromagnetic Heisenberg model on a decorated
hexagonal lattice, involving both classical spins (occupying the vertices) 
and quantum spins (occupying the middle of the links). This study is
motivated by the description 
of a recently synthesized molecular magnetic compound.
First, we trace out the spin $\dem$ degrees of freedom to obtain a 
fully classical model with
an effective ferromagnetic interaction. Then, 
using high temperature expansions and Monte Carlo 
simulations, we analyse its 
thermal and magnetic properties. We show that it provides a good
quantitative description of the magnetic susceptibility 
of the molecular magnet in its paramagnetic phase.
\end{abstract}
The Heisenberg model\cite{Heis} has a long history and has been extensively
studied throughout these last thirty years. While it is exactly solvable in
one dimension\cite{Fisher} in some of its versions, it is only through
approximate methods that quantitative information can be obtained in higher
dimensions. High and low temperature expansions\cite{DombGreen,McKenzie},
Monte Carlo simulations\cite{Binder,Wolf} and renormalisation group
calculations\cite{renorm,BZJ} have been widely developed and give now a
precise account of the critical regime of the model. However, little has
been done in the various specific contexts which are now realized in the
magnetic molecular materials.

For instance, the compound (NBu$_4)_2$Mn$_2$[Cu(opba)]$_3\cdot$6DMSO$\cdot$H%
$_2$O, recently synthesized by H.O. Stumpf {\em et al}\cite{Kahn}, exhibits
a transition at $T_c=15$K towards an ordered state. The structure of this
material can be schematically described by a superposition of layers of
hexagonal lattices with the Mn$^{II}$ ions occupying the vertices 
and the Cu$^{II}$ ions occupying the middle of the links,
as shown in Fig. 1. The interplane-coupling is small, so that
the spin system can be considered two-dimensional. In the plane, the nearest
neighbour Mn-Cu ions interact through an antiferromagnetic coupling. It is
interesting to determine the extent to which such a simple microscopic model
with no other interaction included, can {\em quantitatively} describe the
magnetic and thermal properties of such a complex molecular architecture. Of
course, the isotropic $O(3)$ model is critical only at zero temperature\cite
{MerminW} and the symmetry breaking at $T_c=15K$ has presumably its origin
in a slight spin anisotropy and/or a small interplane coupling. 
However, one
expects for $T\gg T_c$ that the properties of the material are well
described by the two-dimensional isotropic antiferromagnetic spin %
\mbox{$\frac{1}{2}$} - spin \mbox{$\frac{5}{2}$}~interaction. This is the
problem we investigate in this paper.

We denote by ${\bf S}_j^{{\rm (Mn)}}$ the spin $\frac 52$ operator
associated with the Mn ion at site $j$, and by ${\bf S}_i^{{\rm (Cu)}}$ the
spin $\mbox{$\frac{1}{2}$}$ operator corresponding to the Cu ion at site $i$
in the middle of a link of the honeycomb lattice. The antiferromagnetic
interaction is represented by the Heisenberg hamiltonian 
\begin{equation}
{\cal H}=J\sum_{<i,j>}{\bf S}_i^{{\rm (Cu)}}\cdot {\bf S}_j^{{\rm (Mn)}%
}-g_1\mu _B~H~\sum_{j=1}^{N_S}S_j^{z~{\rm (Mn)}}-g_2\mu
_B~H~\sum_{i=1}^{N_L}S_i^{z~{\rm (Cu)}}  \label{Heisenberg}
\end{equation}
where $J$ is positive, $H$ is the external magnetic field, $<i,j>$ stands
for a pair of nearest neighbour spins, $N_S$ is the number of sites and $N_L$
is the number of links on the honeycomb lattice ($N_L=3/2N_S$). The spin $%
\frac 52$ operator can be approximated by a {\em classical} spin $S.{\bf s}$
where ${\bf s}$ is a unit classical vector and $S=\sqrt{\mbox{$\frac{5}{2}$}(%
\mbox{$\frac{5}{2}$}+1)}$, whereas the spin $\mbox{$\frac{1}{2}$}$ operators
are expressed in terms of the Pauli matrices, ${\bf S}^{{\rm (Cu)}}=%
\mbox{$\frac{1}{2}$}~\mbox{\boldmath $\sigma$}$. Since the quantum spin
sites are not directly coupled to each other, one can trace out the quantum
spin dependence to get a completely classical partition function 
\begin{equation}
Z(T,H)=\int {\cal D}\mbox{\boldmath $s$}~\left\{ \prod_{<ij>}2\cosh \left(
\left\| -\mbox{$\frac{1}{2}$}\beta JS(\mbox{\boldmath $s$}_i+\mbox{\boldmath
$s$}_j)+\alpha _2H\hat {{\bf e}}_z\right\| \right) \right\} ~\exp \left(
\alpha _1H\sum_{i=1}^{N_S}s_i^z\right)   \label{Z}
\end{equation}
where we have defined $\alpha _1=S~\beta g_1\mu _B$, 
$\alpha _2=\mbox{$\frac{1}{2}$}\beta g_2\mu_B$,
$\ {{\cal D}\mbox{\boldmath $s$}=\prod_{j=1}^{N_S}\sin \theta
_j~d\theta _j~d\varphi _j }$
and $\Vert {\bf X}\Vert $ stands for the length of vector ${\bf X}$. The
indices $i$ and $j$ now label the {\em classical} spins located at the
vertices of the honeycomb lattice.
\begin{center}
\begin{figure}
\mbox{\epsfig{file=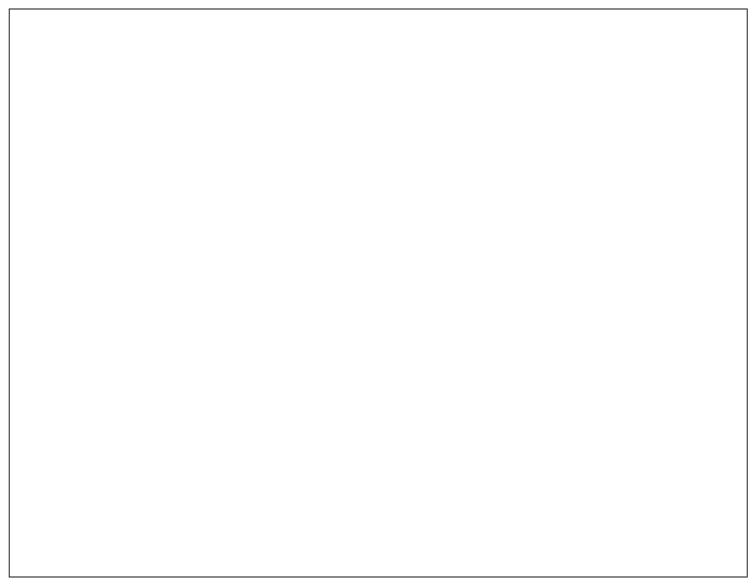}}\\
\caption{Structure of a layer in  
(NBu$_4)_2$Mn$_2$[Cu(opba)]$_3\cdot$6DMSO$\cdot$H$_2$O}
\end{figure}
\end{center}
The partition function of Eq.(\ref{Z}) will now be treated by the standard
techniques of high temperature expansion and Monte Carlo simulation to
extract quantitative information on the system.

In zero magnetic field, the partition function becomes 
\begin{equation}
Z(T,0)=\int {\cal D}\mbox{\boldmath $s$}~\left\{ \prod_{<ij>}2\cosh (%
\mbox{$\frac{1}{2}$}\beta JS\Vert \mbox{\boldmath $s$}_i+\mbox{\boldmath $s$}%
_j\Vert )\right\}   \label{Z0}
\end{equation}
We get an effective {\em ferromagnetic} interaction between the classical
spins~: 
\begin{equation}
-\beta ~{\cal H}_{{\rm eff}}=\sum_{<i,j>}\ln \left[ 2\cosh (%
\mbox{$\frac{1}{2}$}\beta JS\Vert \mbox{\boldmath $s$}_i+\mbox{\boldmath $s$}%
_j\Vert )\right]   \label{Heff}
\end{equation}
We can perform a high temperature expansion of the partition function of Eq.(%
\ref{Z0}). By using the star graph technique\cite{McKenzie}, we have derived
the expansion of $\ln Z(T,0)$ up to the 30th order in the variable $%
\displaystyle{K=\mbox{$\frac{1}{2}$}\beta JS}$. For the specific heat we
obtain the following result~: 
\begin{eqnarray}
{C_V} &=&{N_Lk_B}\left[ 2K^2-\frac{10}3~K^4+4K^6-\frac{8326}{2025}~K^8+\frac{%
3676}{945}~K^{10}-\frac{2963432}{893025}~K^{12}+\frac{43060432}{21049875}%
~K^{14}\right.   \nonumber \\
&&+\frac{1084428794}{1915538625}~K^{16}-\frac{50703530596}{9577693125}%
~K^{18}+\frac{174515087256364}{13540176324675}~K^{20}-\frac{10010372498598008%
}{417635308715625}~K^{22}  \nonumber \\
&&+{\frac{1712584839620191683704}{43895559122555765625}}~K^{24}-{\frac{%
1634086374908287292656}{27958094579597056875}}~K^{26}  \nonumber \\
&&\left. +{\frac{218588272951603892608}{2641543327626328125}}~K^{28}-{\frac{%
9205154548418515452736832}{81777426645321391359375}}~K^{30}\right] 
\label{serieCV}
\end{eqnarray}
According to Eq.(\ref{Z}), the zero-field susceptibility, defined by $%
\displaystyle{\chi =\frac{k_BT}V~\left. \frac{\partial ^2\ln Z}{\partial H^2}%
\right| _{H=0}}$, can be expressed as~: 
\begin{eqnarray}
\chi =\frac{k_BT}V\frac 1{Z(T,0)}\int {\cal D}\mbox{\boldmath $s$}~\left\{
\prod_{<ij>}2\cosh W_{ij}\right\} \left[ \left( \alpha _1\sum_is_i^z+\alpha
_2\sum_{<ij>}\bar s_{ij}^z\right) ^2+\alpha _2^2\sum_{<ij>}Q_{ij}\right] 
\label{chi}
\end{eqnarray}
where $W_{ij}=\mbox{$\frac{1}{2}$}\beta JS\Vert \mbox{\boldmath $s$}_i+%
\mbox{\boldmath $s$}_j\Vert$, $W_{ij}^z=-\mbox{$\frac{1}{2}$}\beta
JS(s_i^z+s_j^z)$,
$\bar s_{ij}^z=\tanh (W_{ij})~W_{ij}^{~z}/W_{ij}$ and
$$
Q_{ij}=\frac{\tanh (W_{ij})}{W_{ij}}~\left[ 1-\left( \frac{W_{ij}^{~z}}
{W_{ij}}\right) ^2\right] +\left( \frac{W_{ij}^{~z}}{W_{ij}}\right) ^2-
(\bar s_{ij}^z)^2
$$

The high temperature expansion of $\chi $ can be obtained in two independent
ways~: first by expanding the partition function of Eq.(\ref{Z}) both in
powers of $K$ and $H$ and retaining the coefficient of $H^2$; second,
through an expansion of the correlation functions which occur in Eq.(\ref
{chi}). Up to the 7th order, we apply both methods in order to validate our
results. By using the second approach, more tractable at higher orders, we
have obtained the following series, up to the order 11~: 
\begin{eqnarray}
T\chi  &=&\frac{N_L}V\frac{\mu _B^2}{k_B}\left[ \frac 29g_{{1}}^2S^2+\frac 14%
g_{{2}}^2-\frac 23g_1g_2SK+\frac 29(g_1^2S^2+g_2^2)K^2\right.   \nonumber \\
&&-\frac 2{27}g_{{1}}g_{{2}}SK^3-{\frac 1{15}}g_{{2}}^2K^4-{\frac 8{405}}%
g_{{1}}g_{{2}}SK^5+\left( {\frac 2{225}}g_{{1}}^2S^2+{\frac{533}{8505}}g_{{2%
}}^2\right) K^6  \nonumber \\
&&+{\frac 2{8505}}g_{{1}}g_{{2}}SK^7-\left( {\frac 4{2835}}g_{{1}}^2S^2+{%
\frac{5683}{127575}}g_{{2}}^2\right) K^8-{\frac 4{4725}}g_{{1}}g_{{2}}SK^9
\nonumber \\
&&\left. +\left( {\frac{524}{893025}}g_{{1}}^2S^2+{\frac{19912}{601425}}g_{{2%
}}^2\right) K^{10}+{\frac{7108}{49116375}}g_{{1}}g_{{2}}SK^{11}\right] 
\label{serieChi}
\end{eqnarray}

In order to improve the range of validity of the expansions (Eqs.(\ref
{serieCV},\ref{serieChi})) we performed a Pad\'e approximant
extrapolation toward low temperatures (large $K$ values). For both series we
get good stability of the Pad\'e table. The results will be presented below.

We performed a self-consistent check of our results by means of a Monte
Carlo simulation of the effective classical model (Eq.(\ref{Heff})). The
various observables can be expressed as ensemble averages with respect to
the Boltzmann weight ${1/{Z(T,0)}e^{-\beta ~{\cal H}_{{\rm %
eff}}}}$. For instance, if we define 
$ {E=-k_BT\sum_{<i,j>}W_{ij}~\tanh (W_{ij})}$
which is the energy of the quantum spins for a fixed configuration
of the classical ones, we find that the internal energy is simply given by $%
\left\langle E\right\rangle _{{\cal H}_{{\rm eff}}}$ and the specific heat
by 
\begin{equation}
C_V=k_B~\beta ^2~\left[ \left\langle E^2\right\rangle _{{\cal H}_{{\rm eff}%
}}-\left\langle E\right\rangle _{{\cal H}_{{\rm eff}}}^2\right]
+k_B\left\langle \Phi \right\rangle _{{\cal H}_{{\rm eff}}} 
\quad \mbox{with}\quad
\Phi =\sum_{<i,j>}\left[ \frac{W_{ij}}{\cosh (W_{ij})}\right] ^2  \label{phi}
 \label{cv}
\end{equation}
Similarly, from Eq.(\ref{chi}) the susceptibility can be expressed as a
sample average. We have used the Wolf algorithm\cite{Wolf} adapted to our
effective Boltzmann weight, on lattices of size increasing with $K$, up to $%
2^{16}$ hexagons for $K=5$.

In Fig. 2 we have plotted the specific heat as a function of $K$. The data
points correspond to the Monte Carlo simulation and the continuous line to
the highest order diagonal Pad\'e approximant of the high temperature
expansion. The first remarkable feature is the well marked knee-hump
variation of $C_V$ as the temperature decreases. One can understand this
effect in the following way. At very low temperature, the system is
dominated by the effective ferromagnetic interaction between the classical
spins. For the purely classical Heisenberg model, one expects a peak at 
low temperature in the specific heat\cite{ital}, corresponding to the 
crossover between the low temperature critical regime and the 
high temperature uncorrelated one.
\begin{center}
\begin{figure}
\mbox{\epsfig{file=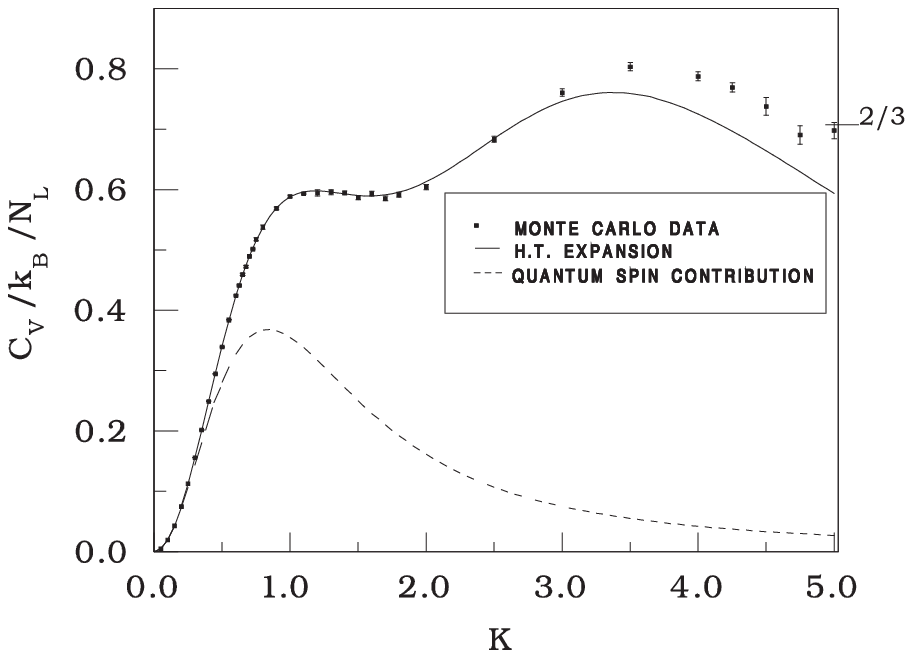}}\\
\caption{ Specific heat $C_V/N_Lk_B$ versus $K=1/2\beta JS$ : The Monte Carlo results are represented by squares, the solid
line shows the Diagonal (7,7) Pad\'e approximant corresponding to the high
temperature series of Eq.(5) and the contribution of the quantum
spins obtained from Eq.(9) is shown as a dashed line. The 
zero-temperature ($K\rightarrow\infty$) limit, $C_V/N_Lk_B=2/3$, 
is indicated. }
\end{figure}
\end{center}
The hump observed in Fig. 2 near $K=4$ corresponds to this
effect. As $T$ increases, the classical spin system goes rapidly to a
disordered state, whereas the quantum spins remain locally coupled to their
classical neighbours. The knee at $K=1$ corresponds to this local
antiferromagnetic order. This effect can be quantitatively confirmed by the
following calculation. Assuming that the classical spins are completely
random leads to $\left\langle E^2\right\rangle =\left\langle E\right\rangle
^2$ in Eq.(\ref{cv}) and 
\begin{equation}
C_v=k_B\left\langle \Phi \right\rangle _{\mbox{\small{disordered}}%
}=8N_Lk_BK^2\int_0^1\frac{y^3}{\cosh ^2(2Ky)}dy  \label{Cvq}
\end{equation}
This function, displayed in Fig. 2 (dashed line), exhibits a peak
exactely under the bump observed in the full $C_v$.

The other feature which emerges from Fig. 2 is the spectacular agreement
of the series expansion with the Monte Carlo results up to $K\simeq 3$ and
the ability of this series to reproduce the double-bump structure. 

We obtain the same perfect agreement of both our methods for the magnetic
susceptibility up to $K=3$. In Fig. 3, we have plotted (solid line) the
(6,6) Pad\'e approximant of the series Eq.(\ref{serieChi}) as a function of
temperature, in a range which corresponds to $0<K<1.2$. The data points
correspond to the experimental results which will be discussed below. The
shape of this curve can be explained as follows. Here again there are two
physically different and competing correlation effects. The first one, due
to the antiferromagnetic spin compensation, leads to a decrease of the
susceptibility, more pronounced at lower temperature. The other contribution, 
due to the ferromagnetic correlation of the classical spins, induces a
divergence at $T=0$. 
\begin{center}
\begin{figure}
\mbox{\epsfig{file=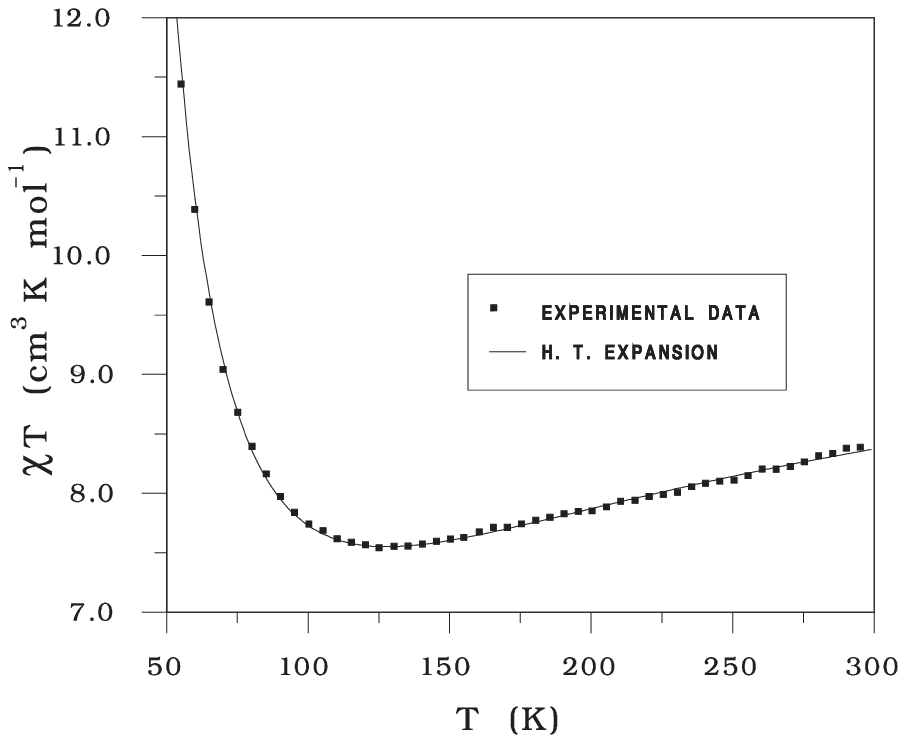}}\\
\caption{Magnetic susceptibility times temperature $\chi T$ in units
of $\mbox{cm}^3:\mbox{K}:\mbox{mol}^{-1}$ versus $T$ in Kelvin: The
experimental data (squares) are taken from $[9]$; the solid line
corresponds to the diagonal (6,6) Pad\'e approximant based on the high
temperature series Eq.(7) with $J=47.6$ K, $g_1=$2.0, $g_2=$2.2.  }
\end{figure}
\end{center}
According to universality, we expect the critical behaviour at zero
temperature to be described by the non-linear $\sigma $-model. At low
temperature the effective interaction Eq.(\ref{Heff}) reduces to 
\[
{\cal H}_{{\rm eff}}\simeq -\mbox{$\frac{1}{2}$}JS\sum_{<ij>}\Vert %
\mbox{\boldmath $s$}_i+\mbox{\boldmath $s$}_j\Vert \approx {\rm const}+\frac 
18JS\sum_{<ij>}\theta _{ij}^2
\]
which corresponds to the low temperature limit of the ordinary classical
Heisenberg model on the honeycomb lattice with the ferromagnetic exchange
constant $J^{*}=\frac 14JS$. The long wave-length expansion leads us further
to the total effective energy in the form of the non-linear $\sigma $-model
Hamiltonian 
\begin{equation}
H_\sigma =\frac{JS}{4\sqrt{3}}~\frac 12\int \left( \partial _\alpha {\bf n}%
\right) \left( \partial _\alpha {\bf n}\right) {\rm d}^2r  \label{NLSM}
\end{equation}
where ${\bf n}({\bf r})$ is the three dimensional unit vector field on the
two dimensional plane ${\bf r}$. As an immediate consequence from the
quadratic expansion we have the magnon contribution to the total energy per
classical spin which is linear in $T$: 
$$\overline{E}\approx -\frac 32JS+k_BT$$
leading to the constant specific heat at low temperature 
$C_V=\frac 23N_Lk_B$. One can see in Fig. 2 that our 
Monte Carlo data are compatible with this value.

As another consequence, we can borrow the low temperature behaviour of the
magnetic susceptibility from the renormalisation group results\cite
{BZJ,CHI_R} for the non-linear $\sigma $-model of Eq.(\ref{NLSM})

\begin{equation}
\chi ={\rm const}\cdot T^3\exp \left[ \frac{4\pi }{k_BT}\frac{JS}{4\sqrt{3}}%
\right] ={\rm const}\cdot T^3\exp \left[ \frac{2\pi }{\sqrt{3}}K\right]
\label{chias}
\end{equation}
\begin{center}
\begin{figure}
\mbox{\epsfig{file=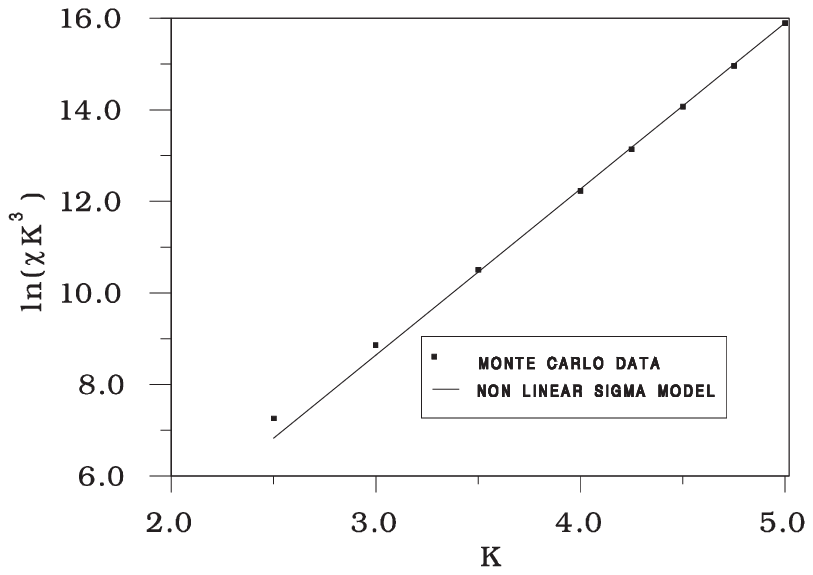}}\\
\caption{$\ln \left( \chi K^3\right) $ versus $K$: The squares
correspond to Monte Carlo results while the solid line shows the prediction
of Eq.(11) of the non-linear $\sigma $-model.}
\end{figure}
\end{center}
In Fig. 4 we have plotted the low temperature Monte Carlo data for $\ln
(K^3\chi )$ as a function of $K$. The straight line corresponds to the
behaviour of Eq.(\ref{chias}). We obtain very good agreement, thereby
confirming that at our lowest temperatures, we are well inside the expected
universality region and the size of our simulated lattices is large enough.%

These results show that our series expansions constitute a reliable
parametrisation of the specific heat and the magnetic susceptibility of the
model up to $K\approx 3$. We can now apply this parametrisation to the
experimental data\cite{Kahn} We have fitted the
product $T\chi $ as a function of the temperature in the range $60K\leq
T\leq 300K$ with $J$, $g_1$ and $g_2$ as free parameters. The result is
shown in Fig. 3, with the best-fit parameters given by $J=47.6$K, $g_1=2.0$
and $g_2=2.2$. We observe excellent agreement with the experimental data,
confirming that the high temperature magnetic properties of this compound
are well described by our model. Furthermore, these values of the parameters
are very close to those obtained for both Cu-Mn pairs\cite{Kahn2} and 
chains\cite{Kahn,Georges} with the same bridging network.
The weakly pronounced minimum observed in Fig. 3 for $T\approx 120$K
has its origin in the local antiferromagnetic ordering discussed above.

The material exhibits a ferromagnetic transition at $T_c=15$K that the
isotropic model cannot describe. To give an account of this
nonzero-temperature critical behaviour the Hamiltonian must be generalized
to include spin anisotropy and three-dimensional effects\cite{Anis}, a
modification which we are now investigating. However, it is amazing that the
agreement between the experimental data and the model persists down to
rather low temperatures ($T\approx 20$K), close to the measured $T_c$.
\\[0.5cm]

\noindent{{\bf Aknowledgements}}\\
\noindent S.V. M. thanks V.A. Fateev and Al.B. Zamolodchikov for helpful
discussions.


\end{document}